\documentclass[useAMS,usenatbib]{mn2e}

\usepackage{graphics}
\usepackage{graphicx}
\usepackage{rotating}
\usepackage{lscape}
\usepackage{amsmath}
\usepackage{amssymb}

\newcommand{\tobs}{T_\rmn{obs}}
\newcommand{\tol}{T_\rmn{overlap}}

\newcommand{\asqzeta}{A^2_{ij} \zeta\left(\theta_{ij}\right)}
\newcommand{\easqzeta}{\delta\asqzeta}

\newcommand{\e}{\times 10^}

\title[On Detection of the Gravitational-Wave Background]{On detection of the stochastic gravitational-wave background using the Parkes pulsar timing array}
\author[Yardley et al.]{D. R. B. Yardley$^{1,2}$\thanks{E-mail: dyardley@physics.usyd.edu.au (DRBY)}, W. A. Coles$^3$, G. B. Hobbs$^1$, J. P. W. Verbiest$^4$, \newauthor R. N. Manchester$^1$, W. van Straten$^5$, F. A. Jenet$^6$, M. Bailes$^5$, N. D. R. Bhat$^5$, \newauthor S. Burke-Spolaor$^{1,5}$, D. J. Champion$^{4,1}$,  A. W. Hotan$^7$, S. Oslowski$^{1,5}$, \newauthor J. E. Reynolds$^1$, J. M. Sarkissian$^1$\\
$^1$CSIRO Astronomy and Space Science, Australia Telescope National Facility, P.O. Box 76, Epping, NSW 1710, Australia.\\
$^2$Sydney Institute for Astronomy, School of Physics A29, The University of Sydney, NSW 2006, Australia.\\
$^3$Electrical and Computer Engineering, University of California at San Diego, La Jolla, California, U.S.A.\\
$^4$Max-Planck-Institut f\"ur Radioastronomie, Auf dem H\"ugel 69, 53121, Bonn, Germany.\\
$^5$Centre for Astrophysics and Supercomputing, Swinburne University of Technology, H31 P.O. Box 218, Hawthorn, VIC 3122, Australia. \\
$^6$Center for Gravitational Wave Astronomy, University of Texas at Brownsville, 80 Fort Brown, Brownsville, TX 78520, USA. \\
$^7$Department of Imaging and Applied Physics, Curtin University, Bentley, Western Australia, Australia. \\
}
\begin{document}
\date{Accepted 2011 February 9.  Received 2011 February 8; in original form 2010 December 18}

\pagerange{\pageref{firstpage}--\pageref{lastpage}} \pubyear{2010}

\maketitle

\label{firstpage}

\begin{abstract}
We search for the signature of an isotropic stochastic gravitational-wave background in pulsar timing observations using a frequency-domain correlation technique. These observations, which span roughly 12\,yr, were obtained with the 64-m Parkes radio telescope augmented by public domain observations from the Arecibo Observatory. A wide range of signal processing issues unique to pulsar timing and not previously presented in the literature are discussed. These include the effects of quadratic removal, irregular sampling, and variable errors which exacerbate the spectral leakage inherent in estimating the steep red spectrum of the gravitational-wave background. These observations are found to be consistent with the null hypothesis, that no gravitational-wave background is present, with 76 percent confidence. We show that the detection statistic is dominated by the contributions of only a few pulsars because of the inhomogeneity of this data set. The issues of detecting the signature of a gravitational-wave background with future observations are discussed.
\end{abstract}


\section{Introduction} \label{sec:intro}

Times-of-arrival (ToAs) of high signal-to-noise (S/N) ratio integrated pulses from millisecond pulsars (MSPs) can be measured very precisely, often with sub-$\mu$s uncertainties. The rotational stability of a MSP implies that a simple model of the pulsar can be developed to make accurate predictions of these ToAs. Comparing the measured ToAs with these predictions enables the study of many astrophysical phenomena; for example, this process led to evidence for the existence of gravitational waves \citep[GWs;][]{tw82}. As more MSPs are discovered and instrumentation is improved, it is becoming likely that pulsar observations will lead to the direct detection of GWs, using their effect on ToAs described independently by \citet{saz78} and \citet{det79}. A GW will cause a perturbation in the ToA when it passes the pulsar and again when it passes the Earth. The perturbations that would be detectable with pulsar timing are expected to have amplitudes of $\sim$10\,ns and timescales greater than one year \citep[see, e.g.,][]{svv09}.

The Parkes Pulsar Timing Array (PPTA) project \citep[e.g.,][]{2010CQGra..27h4015V} is using the Parkes radio telescope and advanced instrumentation to time 20 MSPs over a period of at least 5 years. With careful calibration and long integrations, the majority of the pulsars are yielding weighted root-mean-square (rms) residuals below 1\,$\mu$s, with a few below 200\,ns \citep{2010arXiv1004.3602M}. While some of the PPTA pulsars do show ``timing noise'' (low-frequency timing instabilities which are unmodelled by conventional analyses), \citet{vbc+09} showed that this will not prohibit GW detection with the PPTA pulsars. The project will allow examination of correlated signals between the different pulsars, including detecting variations in the terrestrial timescale \citep[e.g.,][]{pt96,2010arXiv1011.5285H}, detecting errors in the Solar-System ephemeris \citep{2010ApJ...720L.201C}, and providing constraining limits on, or a detection of, low-frequency GWs. The project has been ongoing since late 2004. Observations of some of the PPTA pulsars have been made at the Parkes observatory since 1994, albeit with less regularity and precision.

Recent work \citep{2010MNRAS.407..669Y, 2010arXiv1005.5163B, 2010MNRAS.401.2372V, svv09, 2010arXiv1008.1782C} has addressed the detectability of individual sources of GWs in pulsar timing residuals and shows that it is unlikely that current instrumentation will allow a detection. However, if the universe contains many such sources of GWs, these sources will form an isotropic stochastic gravitational-wave background (GWB). \citet{saz78}, \citet{det79}, \citet{hd83} and \citet{jhlm05} have described how pulsar timing arrays (PTAs), such as the PPTA, can directly detect such a background of $\sim$nHz frequency GWs. For each pulsar, this GWB would cause ToA perturbations that are correlated between pulsar pairs in a quadrupolar fashion. This correlation, which depends only on the angle between the pair of pulsars as shown in Figure 1 \citep{hd83}, provides an unambiguous signature of the GWB. The functional form of this signature is given by:
\begin{equation}
\label{eq:zeta}
\zeta(\theta_{ij}) = \frac{3}{2}x\log x - \frac{x}{4} + \frac{1}{2} \;,
\end{equation}
where $x=[1-\cos(\theta_{ij})] / 2$ and $\theta_{ij}$ is the angle between pulsars $i$ and $j$ subtended at the observer \citep{hd83, jhlm05}. The function $\zeta(\theta_{ij})$ is independent of GW frequency, and is derived assuming GWs are described by general relativity; other GW modes are analysed in \citet{2008ApJ...685.1304L} but are not considered in this paper. We believe that a first detection of the GWB is only possible via an unambiguous detection of this expected correlation. In view of the widespread interest in such a detection, we have designed a detection procedure that can show this signature in an easily discernible and convincing manner.

\begin{figure}
\includegraphics[width=0.33\textwidth, angle = 270]{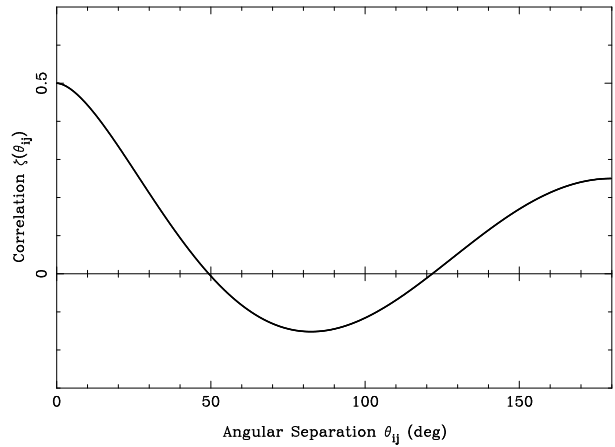}
\vspace{5pt}
\caption{
The expected correlation in pulsar timing residuals due to an isotropic stochastic GWB. The abscissa gives the angle subtended at the observer by a particular pulsar pair. The ordinate gives the expected correlation between the timing residuals of that pair. This signal is independent of the GW frequency and assumes that GWs behave as predicted by general relativity.
}
\label{fig:HD}
\end{figure}

Several techniques have already been proposed in the literature to both limit \citep{1983ApJ...265L..35R, ktr94, 1996PhRvD..53.3468T, lom02, jhv+06} and detect \citep{jhlm05, 2009PhRvD..79h4030A} the GWB. However, these methods have not taken into account all the details of optimally treating pulsar timing data, or are restricted to particular observations. Application of a Bayesian technique \citep{2009MNRAS.395.1005V} is ongoing work, and our method is completely independent. The lowest published limit for a GWB caused by supermassive binary black holes \citep{jhv+06} begins to constrain the parameters of galaxy evolution \citep[e.g.,][]{2003ApJ...590..691W}, cosmic strings \citep[e.g.,][]{dv05} and relic GWs from the Big Bang \citep[e.g.,][]{mag00}. Further improvements in sensitivity could either enable detection of GWs or rule out most proposed models of a GWB.

The GWB detection technique we present here is based on the method of \citet{jhlm05}. It improves on their technique in a number of ways: 
\begin{itemize}
    \item[-] we study the pairwise correlation described by \citet{hd83} in the form of pairwise cross power spectra;
    \item[-] we obtain independent estimates of the GWB from each frequency component in each cross power spectrum;
    \item[-] we use an optimally weighted linear combination of the cross power estimates as the detection statistic;
    \item[-] we account for the effect of different overlapping timespans between the pulsar pairs;
    \item[-] we calibrate the cross power spectra and their estimated errors using simulations that completely account for the fitting of the pulsar timing model.
\end{itemize}

Our technique is not optimal for bounding the GWB with these observations because the variation in S/N ratio between pulsars is too large. This means that there are not enough significant cross power spectra to compensate for the low value of the average cross correlation. A tighter bound on the GWB amplitude could be obtained with these observations using the amplitudes of the individual power spectra \citep[similar to][]{jhv+06}. However, a detection algorithm cannot be based on the amplitudes of individual power spectra because there are many unknown contributions to those power spectra. We discuss a number of issues that are common to both the \citet{jhv+06} limit technique and any limit technique based on measuring the GWB-induced correlation between pulsars. Such issues include the estimation of power spectra when the sampling is irregular and the ToA uncertainties are variable, and the effects of fitting the timing model.

In \S \ref{sec:obsns} we describe the observations and the analysis that led to the timing residuals we use in this paper. \S \ref{sec:method} describes the theoretical background and our method for making a detection of the isotropic stochastic GWB. \S \ref{sec:res} describes the results obtained, \S \ref{sec:disc} describes their implications and the outstanding issues for GWB detection via pulsar timing, and \S \ref{sec:conc} summarises our conclusions.

\section{Observations} \label{sec:obsns}

\begin{table*}
 \centering
 \begin{minipage}{180mm}
  \caption{Basic information for the Verbiest et al. (2008, 2009) data sets.} \label{tbl:jorisData}
  \begin{tabular}{@{}ccccccc@{}}
  \hline
    PSRJ & Period & DM & $P_b$ & Weighted RMS & Span & No. of                 \\
    & (ms) & (cm$^{-3}$pc) & (d) &  Residual ($\mu$s) & (years) & Observations\\
 \hline
J0437$-$4715 & 5.757 & 2.65 		& 5.74 	& 0.20 & 9.9 & 2847    \\
J0613$-$0200 & 3.062 & 38.8 		& 1.20 	& 1.52* & 8.2 & 190    \\
J0711$-$6830 & 5.491 & 18.4 		& -- 		& 3.24* & 14.2 & 227     \\
J1022+1001 & 16.45 & 10.3 		& 7.81 	& 1.63* & 5.1 & 260     \\
J1024$-$0719 & 5.162 & 6.49 		& -- 		& 4.17* & 12.1 & 269      \\
J1045$-$4509 & 7.474 & 58.2 		& 4.08 	& 6.80* & 14.1 & 375*   \\
J1600$-$3053 & 3.598 & 52.2 		& 14.3 	& 1.11* & 6.8 & 474*   \\
J1603$-$7202 & 14.84 & 38.1 		& 6.31 	& 1.98* & 12.4 & 212  \\
J1643$-$1224 & 4.622 & 62.4 		& 147 	& 1.94* & 14.0 & 241 \\
J1713+0747 & 4.570 & 16.0 		& 67.8 	& 0.20 & 14.0 & 392    \\
J1730$-$2304 & 8.123 & 9.61 		& -- 		& 2.52* & 14.0 & 180      \\
J1732$-$5049 & 5.313 & 56.8 		& 5.26 	& 3.23* & 6.8 & 129    \\
J1744$-$1134 & 4.075 & 3.14 		& -- 		& 0.62 & 13.2 & 342      \\
J1824$-$2452 & 3.054 & 120 		& -- 		& 1.63* & 2.8 & 89      \\
J1857+0943 & 5.362 & 13.3 		& 12.3 	& 1.14* & 22.2\footnote{There is a gap of $\sim$11\,years between the end of the observations presented in \citet{ktr94} and the beginning of observations with the Parkes telescope. In our analysis we use the Arecibo observations of PSR J1857+0943 only to assist in the estimation of the pulsar parameters and then discard the Arecibo residuals in further processing.} & 376 \\
J1909$-$3744 & 2.947 & 10.4 		& 1.53 	& 0.17 & 5.2 & 893\\
J1939+2134 & 1.558 & 71.0 		& -- 		& 15.0\footnote{We have altered the value of the phase offsets between different observing systems for these timing residuals compared with the analysis of \citet{vbc+09}, which lowers our measured rms.} & 23.3\footnote{This time series features several large gaps and includes the \citet{ktr94} data.} & 588\\
J2124$-$3358 & 4.931 & 4.62 		& -- 		& 4.01* & 13.8 & 415*   \\
J2129$-$5721 & 3.726 & 31.9 		& 6.63 	& 2.19 & 12.5 & 177*\\
J2145$-$0750 & 16.05 & 9.00 		& 6.84 	& 1.88* & 13.8 & 376*\\
\hline
\end{tabular}
\footnotetext[0]{\hspace{-0.18cm}*\hspace{0.08cm} These values differ slightly from those presented in \citet{vbc+09} because we have removed duplicated observations in five pulsars, and corrected a minor processing error involving the uncertainties on observations made with different observing systems.}
\vspace{-0.5cm}
\end{minipage}
\end{table*}

\begin{figure}
\includegraphics[width=0.28\textwidth, angle = 270]{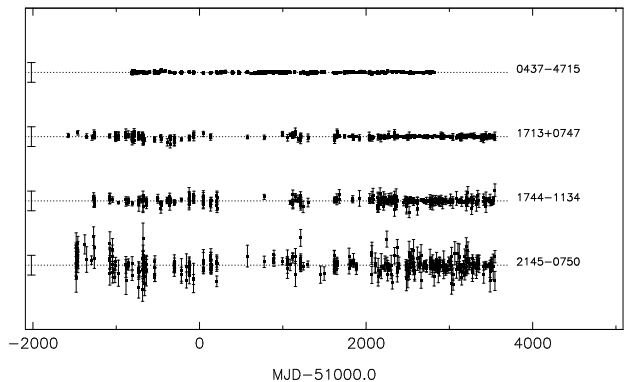}
\vspace{5pt}
\caption{
The timing residuals for the four most influential pulsars used in this analysis. The length of the vertical line on the left-hand side indicates 10\,$\mu$s. The right-hand column gives the pulsar's J-name. Noise levels vary significantly both between pulsars and at different epochs.
}
\label{fig:overlaps}
\end{figure}

The 20 pulsars used in this paper were observed for $\sim$10\, min to 1\,h in each observation, depending on the hardware being used at the time. Since 2005, the typical integration time on most pulsars is $\sim$ 1\,h. For each observation a mean pulse profile was formed using an ephemeris which ``folds'' the data at the apparent pulse period. Observations of each pulsar were made every few weeks (although there are some gaps of many months) and the observations span many years, as shown in Table \ref{tbl:jorisData} and Figure \ref{fig:overlaps}. The time shift between a standard pulse profile and the observed profile is measured using the technique described in \citet{tay92}, as implemented within the \textsc{pat} routine of the \textsc{PSRCHIVE} software package \citep{hvm04,2010PASA...27..104V}. This measurement results in an estimate of a ToA and its uncertainty. The observatory timescale was referenced to Universal Coordinated Time and post-corrected to Terrestrial Time as realised by International Atomic Time, abbreviated to TT(TAI). The effect of corrections to this timescale published by the Bureau International des Poids et Mesures (BIPM) is discussed in more detail in \S\ref{sec:corrsig}. The ToAs were transformed to a barycentric arrival time using the DE405 Solar-System ephemeris \citep{sta04b}. The barycentric ToAs were then fitted with a timing model using the \textsc{tempo2} software package \citep{hem06, ehm06}. We refer to the differences between the observed and predicted ToAs as the ``timing residuals''. Statistically significant timing residuals represent physical effects that have not been included in the timing model, and can have many causes, such as incomplete polarisation calibration, timing noise intrinsic to the pulsar system, fluctuations in the ISM, errors in the Solar-System ephemeris, errors in the terrestrial timescale and GWs. More on the techniques of pulsar timing can be found in \citet{2004hpa..book.....L} and \citet{ehm06}.

In this paper we use the timing residuals presented by \citet{vbv+08} and \citet{vbc+09}. These residuals are assembled from a large number of different observations with different receivers and even different observatories. The observations come primarily from the Parkes radio telescope and most were made as part of the PPTA project. They are augmented by earlier Parkes observations and publicly available observations of PSR J1857+0943 and PSR J1939+2134 taken with the Arecibo radio telescope and described in \cite{ktr94}. The Arecibo observations of PSR J1857+0943 were carried out at $\sim$1400\,MHz and span seven years. The Arecibo observations of PSR J1939+2134 were carried out at $\sim$1400\,MHz and $\sim$2400\,MHz and span eight years.

The Verbiest et al. (2008, 2009) observations were performed in the 20\,cm (1400\,MHz) band, except for PSR J0613$-$0200 for which a better timing solution was obtained in the 50\,cm (685\,MHz) band. The observations have not been fully corrected for variations in the pulse dispersion measure (DM). Observations in the 20cm band between 1994 and November 2002 were made with either one or two 128\,MHz-wide bands, but these data vary greatly in quality. Observations after November 2002 were taken over two 64\,MHz-wide observing bands centred at 1341\,MHz and 1405\,MHz. For full details of ToA measurement and data processing, see \citet{vbc+09}. As mentioned in the footnote to Table \ref{tbl:jorisData}, we have made minor corrections to the Verbiest et al. (2008, 2009) observations, though we have not performed the full data reduction process already described and performed in Verbiest et al. (2008, 2009). We have also altered the value of the phase offsets between different observing systems for the PSR J1939+2134 timing residuals, which reduces the measured rms. A summary of the data sets is given in Table \ref{tbl:jorisData}. Of the 20 time series, the four most influential are plotted in Figure \ref{fig:overlaps}; the timing residuals from all observations are shown in Figure 1 of \citet{vbc+09}.

The observations were made with a number of different observing systems $-$ both the frontend receivers and the backend instrumentation have varied over time. Arbitrary phase offsets have been fitted for and removed between the ToAs from each different system for a given pulsar. This reduces the noise level in the timing residuals for that pulsar, especially over long timescales. These residuals also have a number of features which complicate the time series analysis and spectral estimation. While the timing residuals of most of the pulsars are ``white'' (i.e., their power spectra are independent of frequency), nine out of the twenty pulsars exhibit non-white noise. This was determined using a simple two-point correlation analysis to determine the degree of correlation between adjacent residuals using the \textsc{checkwhite} plugin to \textsc{tempo2}.

The data spans vary widely, ranging from 2.8 years for PSR J1824$-$2452 to 23.3 years for PSR J1939+2134. The weighted rms residual also varies over two orders of magnitude, from 170\,ns for PSR J1909$-$3744 to 15\,$\mu$s for PSR J1939+2134. The residuals are also sampled irregularly and the sampling is different between pulsars. The ToA uncertainties for a given pulsar vary widely over short and long timescales. This is normally caused by scintillation in the interstellar medium and upgrades in the receiver and backend systems, respectively. In some cases, the magnitude of the ToA error bar changes discontinuously during the time series due to these upgrades in the observing hardware at Parkes. The upgrade which had the largest effect on the quality of the timing residuals was the transition from the Caltech incoherent autocorrelation spectrometer fast pulsar timing machine \citep{1994PhDT........12N} to the Caltech-Parkes-Swinburne Recorder 2 \citep{2003ASPC..302...57B}, a coherent dedispersion system, in late 2002. We therefore attempt to reduce the huge variation in the magnitude of the ToA uncertainties so that, in subsequent weighted estimates using the timing residuals, the weights are spread more evenly across the data set. We provide in Table \ref{tbl:nonstat} a list of the pulsars for which we have calculated the sample variance of the residuals in two different sections of the time series because of a step-change in the quality of the timing residuals. These sample variances are added in quadrature with the original error bars in each portion before commencing any further processing. For all other pulsars there was no significant change in data quality at the epoch of the hardware change. We thus calculate the sample variance of the whole time series and add it in quadrature with the original error bars before any further processing.

\begin{table}
  \caption{Pulsars with non-stationary timing residuals. For these pulsars, we estimate the unweighted rms of the residuals before and after an important hardware change at the telescope.} \label{tbl:nonstat}
  \begin{tabular}{@{}ccccc@{}}
  \hline
PSRJ & Type of 	& Epoch   		&  RMS before	  	&  RMS after \\
    	  & change	& (MJD)		&  change ($\mu$s)  	&  change ($\mu$s)\\
 \hline
J1600$-$3053 	& backend & 52654.0 & 9.61 & 1.31 \\
J1713+0747 	& backend & 52462.5 & 1.24 & 0.48 \\
J1732$-$5049 	& backend & 52967.5 & 7.57 & 4.03 \\
J1744$-$1134 & backend & 52462.6 & 1.54 & 1.29 \\
J2124$-$3358 & backend & 52984.5 & 9.74 & 4.64 \\
J2129$-$5721 & receiver & 51410.0 & 5.47 & 3.48 \\
J2145$-$0750 & backend & 52975.5 & 4.14 & 3.17 \\
 \hline
\end{tabular}
\end{table}

\section{Method} \label{sec:method}

For all pulsars, the GWB will induce timing residuals with a steep red power spectrum. These induced residuals are correlated between different pulsar pairs as shown in Figure \ref{fig:HD}. Although limits on the amplitude of the GWB can be obtained from the residuals of a single pulsar \citep[see, e.g.,][]{ktr94}, the GWB can only be detected with confidence by observing this pair-wise correlation. In pulsar timing residuals, ``red'' (i.e., low-frequency) power can come from a variety of other physical effects. These include irregular spindown behaviour known as ``timing noise'' \citep[][and references therein]{2010MNRAS.402.1027H,2010ApJ...725.1607S}, variation in the pulse dispersion in the interstellar medium \citep{yhc+07} or calibration and other instrumental errors \citep{van06}. There are also some sources of noise which are correlated between pulsars, such as instabilities in Terrestrial Time (TT) and inaccuracies in the Solar-System ephemeris \citep{2010ApJ...720L.201C}. An instability in TT will affect all pulsars in the same way, inducing a correlated signal which is independent of the angular separation of the pulsars on the sky, leading to a positive offset in the correlation curve in Figure \ref{fig:HD}. An inaccuracy in the Solar-System ephemeris will induce residuals which are positively correlated for pairwise angular separations less than 90 degrees. Such a signal could be correlated with the GWB signal shown in Figure \ref{fig:HD}. All of these low-frequency variations are difficult to predict and need to be accounted for when implementing an algorithm to detect the GWB.

\subsection{The Expected GWB Signal}

Throughout this paper we assume a power-law form for the characteristic strain, $h_c(f)$, of the isotropic stochastic GWB. This power-law is given by \citep[e.g.,][]{2001astro.ph..8028P, jhv+06}
\begin{equation}
\label{eq:h_c}
h_c(f) = A\left(f / f_{1\,\rmn{yr}}\right)^{\alpha}
\end{equation}
where $f$ is the GW frequency, $f_{1\,\rmn{yr}}=(1\,\rmn{yr})^{-1}$ and $A$ is a dimensionless quantity termed the ``amplitude'' of the GWB. The smallest upper bound on the amplitude of the GWB from the literature is $A \leq 1.1 \times 10^{-14}$ \citep{jhv+06}. The power-law form of the GWB is consistent with most models to date \citep[e.g.,][]{jb03, 2003ApJ...590..691W}. The spectral exponent, $\alpha$, can take a range of values depending on the source of the GWB under investigation (e.g., cosmic strings, small-orbit black-hole binaries). However, all predicted backgrounds have $\alpha < 0$ \citep[][and references therein]{jhv+06}, which results in a steeply decreasing power spectrum in the timing residuals. Several models of the expected GWB from an ensemble of supermassive black-hole binaries (SMBHBs) predict that the amplitude of the GWB will be in the range $5\times10^{-16} < A < 10^{-14}$ \citep{jb03, 2003ApJ...590..691W, svc08} with a spectral exponent $\alpha = -2/3$.\footnote{\cite{svc08} proposed a more complicated frequency-dependence for $h_c(f)$ involving an extra term proportional to $f^{-1}$, which causes significant deviation from equation (\ref{eq:h_c}) for $f > 10^{-8}$\,Hz; current PTA projects are not yet sensitive enough to distinguish between the two forms. Until a detection of the GWB is made, PTAs are expected to focus on frequencies $f \ll 10^{-8}$\,Hz.}

If the GWB strain is described by equation (\ref{eq:h_c}), then the power spectrum of the induced ToA perturbations is \citep[see, e.g.,][]{det79, jhlm05, jhv+06}
\begin{equation}
\label{eq:P_g}
P_g(f) = \frac{A^2}{12\pi^2} \left(\frac{f}{f_{1\,\rmn{yr}}}\right)^{2\alpha - 3}.
\end{equation}
The cross power spectrum between the induced ToA perturbations in pulsars $i$ and $j$ is
\begin{equation}
\label{eq:X_ij}
X_{ij}(f) = P_g(f) \zeta\left(\theta_{ij}\right)
\end{equation}
where $\zeta\left(\theta_{ij}\right)$ is given in equation \ref{eq:zeta}.

\subsection{Detecting the GWB signal} \label{sec:det}

We estimate $X_{ij}(f)$ for each pair of pulsars. As the spectrum of the GWB is very steep, only the lowest frequencies are of interest. Fortunately, the irregular sampling has less effect on the lower frequencies than on the higher frequencies because the low frequencies are heavily oversampled. The observations of each pair of pulsars overlap over some time span $\tol$. For $N_{\rm psr} = 20$ there are $N_{\rm pairs} = 190$ pairs. For each pair we estimate the cross power spectrum at harmonics of $f = 1/\tol$. If the sampling were uniform, these estimates would be uncorrelated. In practice we find that they are not uncorrelated and this reduces the sensitivity of our detection algorithm. It is probable that the independence can be restored using the Cholesky spectral estimation procedure recently discussed by \citet{2010Coles}. However, this is beyond the scope of this work.

For some pairs, $\tol$ can be much smaller than the length of one or both time series. For our time series, $\tol$ ranges from just 0.8\,yr for PSRs J0437$-$4715 and J1824$-$2452, to 14.1\,yr for PSRs J0711$-$6830 and J1939+2134. The use of only the overlapping residuals causes a bias in the cross power spectral estimates, the causes of which are currently not known. We correct this bias by removing a quadratic function from the overlapping section of the two time series using a weighted least-squares (WLSQ) fit, as shown in Figure \ref{fig:fit}. This fit is in addition to the standard timing model fit which estimates the pulsar parameters. We estimate the cross power spectrum:
\begin{equation}
\label{eq:cs}
X_{ij}(f) = \mathcal{F}_i(f) \mathcal{F}_j^*(f) / \tol
\end{equation}
where $\mathcal{F}_i$ denotes the Discrete Fourier Transform (DFT) of the timing residuals of pulsar $i$ and $^*$ denotes complex conjugation. We use the following standard definition of the one-sided DFT:
\begin{equation}
\label{eq:dft}
\mathcal{F}(f_k) = 2 \sum_{n=0}^{N-1} \frac{r_n}{N} e^{-2\pi i k n} ,
\end{equation}
where $i = \sqrt{-1}$ in this particular case, $N$ is the number of timing residuals, $r_n$ is the $n$-th residual and $k$ is an integer between 1 and $(N - 1) / 2$, rounded down. Note that the $k=0$ term corresponds to the mean of the time series, which is zero for pulsar timing residuals. Calculating the DFT is not trivial because of the uneven sampling and variable error bars. We calculated $\mathcal{F}_i(f_k)$ for every pulsar using a WLSQ fit of a sine term plus a cosine term at each $f_k = k / T_{\rm overlap}$. This gives identical results to a weighted Lomb-Scargle estimate of the spectrum \citep{sca82,2009A&A...496..577Z}. The variance of each cross-power spectral estimate is
\begin{equation}
\label{eq:delpij}
\sigma^2_{X_{ij}}(f) = \langle P_i(f)\rangle \langle P_j(f)\rangle  /  2  
\end{equation}
where $\langle ... \rangle$ indicates an expectation value and $P_i(f)$ is the spectral estimate of the residuals of pulsar $i$ at frequency $f$. In practice, we calculate these expectation values using a power-law fit to the lowest frequencies in the spectrum of each pulsar. This power-law fit gives a spectral model for low frequencies in this pulsar.

We account for the effects of fitting the timing model to the observations using two Monte Carlo simulations. The first simulation estimates the power spectrum $-$ before and after pulsar parameter fitting $-$ of simulated white noise with the same sampling and ToA errors as the residuals of each pulsar. Dividing the post-fit power spectrum by the pre-fit power spectrum gives the effective ``transfer function'' of the full \textsc{tempo2} fitting procedure \citep[see, e.g.,][]{bnr84,1989NASCP3046...93H}, and this process is repeated 1000 times to find the average transfer function. We describe this as an \emph{effective} transfer function because the \textsc{tempo2} fitting process does not act exactly as a filter. We correct the measured cross power spectrum for each pulsar pair at each frequency by dividing by the geometric mean of the transfer functions of the two pulsars at that frequency. This correction is common between our analysis and that of \citet{vbc+09}, but this is the only pulsar parameter fitting correction that Verbiest et al. perform.

However, the transfer function can only correct the effects of the timing model fit as it acts on white noise in the residuals because, although fitting the timing model is a linear operation, it is not a filter. When the residuals are affected by red noise, fitting the full timing model to the residuals reduces $P(f=1/\tobs)$ by considerably more than the white noise transfer function, where $\tobs$ is the time span of the residuals. This is easily shown by simulation. A second correction is therefore necessary to measure the effect of the full timing model fit on the non-white GWB contribution to the residuals. We simulate $\sim$10000 realisations of the residuals and add a simulated GWB signal with $A = 3\e{-15}$ and $\alpha = -2/3$ to all pulsars using the method described in \citet{hjl+09}. This value of $A$ was chosen because it gives the largest GWB signal which is still small compared with the noise, hence reducing the number of required simulations. We further reduce the number of simulations by fixing every pulsar to be at the same position and distance, giving the maximum correlated GWB signal between pulsars. We perform the full pulsar parameter fit using \textsc{tempo2}, estimate the cross power spectrum in each realisation and apply the transfer function correction described above. We divide the average corrected cross power spectrum of each pulsar pair by the theoretical level of the cross power spectrum given in equation (\ref{eq:X_ij}). This process defines a set of ``calibration factors'', $\gamma_{ij}(f_k)$. When forming subsequent estimates of the cross power spectrum using equation (\ref{eq:cs}), we calibrate each estimate at the lowest three frequencies of the cross power spectrum by dividing the cross-power-spectral estimate at frequency $f_k$ for pulsars $i$ and $j$ by $\gamma_{ij}(f_k)$.

After performing both of these corrections, we estimate $A^2$. For each frequency channel, $f_k$, of the cross power spectrum (measured in yr$^{-1}$), we have (cf. Equations \ref{eq:P_g} and \ref{eq:X_ij})
\begin{equation}
\label{eq:a2zeta}
\left[A^2_{ij} \zeta\left(\theta_{ij}\right)\right]_k = 12 \pi^2 f_k^{3-2\alpha} \rmn{Real}\left[X_{ij}(f_k)\right]
\end{equation}
where $A^2_{ij}$ indicates the measurement of $A^2$ obtained from pulsars $i$ and $j$ and $\rmn{Real}\left[X_{ij}(f_k)\right]$ is the real part of the cross power spectrum. The variance of $A^2_{ij} \zeta\left(\theta_{ij}\right)$ is then proportional to the variance of $X_{ij}$.

To compare directly with the technique of \citet{jhlm05}, we perform a weighted sum of the $\asqzeta$ estimates over cross-spectral frequency to obtain a single estimate of $\asqzeta$ for each pulsar pair. 
\begin{equation}
\label{eq:asqz}
\asqzeta = \frac{ 12 \pi^2 \sum_k X_{ij}(f_k) k^{2\alpha - 3} / \sigma^2_{X_{ij}}(f_k) }{ \left(\tol\right)^{3 - 2\alpha}  \sum_k k^{4\alpha - 6} / \sigma^2_{X_{ij}}(f_k)}
\end{equation}
where both summations range from $k=1$ to $N_{{\rm spec}, ij}$, where $N_{{\rm spec}, ij}$ is the number of cross-spectral frequencies for pulsars $i$ and $j$. This final estimate of $\asqzeta$ is similar to the unnormalised covariance between the residuals of pulsars $i$ and $j$. We also use the observed scatter in estimates of $\asqzeta$ obtained from simulated observations to estimate the uncertainty $\easqzeta$ for each pulsar pair.

Having fully calibrated our technique using simulations, we estimate the squared amplitude of the GWB, $\hat{A^2}$, by forming an average of the $\asqzeta$ estimates weighted by the inverse variance of each estimate. In practice this average is done by performing a WLSQ fit to find the amplitude $\hat{A^2}$ (and its corresponding uncertainty) for which the quantity $\hat{A^2} \zeta$ best fits the observed values of $\asqzeta$. For ease of notation, we index over all possible pulsar pairs using $m$, where $m$ is an index running from 1 to $N_{\rm pairs}$ and we set $\zeta_m\equiv\zeta\left(\theta_{ij}\right)$. In this case, the expression for $\hat{A^2}$ is
\begin{equation}
\label{eq:a2est}
\hat{A^2} = \frac{ \sum_m \left[A^2_{m} \zeta_m\right] \zeta_m / \sigma^2_{A^2_{m} \zeta_m}}   {\sum_m \zeta_m^2 / \sigma^2_{A^2_{m} \zeta_m}} =  \frac{ \sum_m A^2_{m} / \sigma^2_{A^2_{m}} }   {\sum_m 1/ \sigma^2_{A^2_{m}}}
\end{equation}
and its unweighted variance is
\begin{equation}
\label{eq:a2esterr}
\sigma_{\hat{A^2}}^{2} = \frac{1}{ \sum_m \zeta_m^2 / \sigma^2_{A^2_{m} \zeta_m}} = \frac{1}{\sum_m 1/\sigma^2_{A^2_m}}   \;\cdot
\end{equation}

This initial estimate of the error assumes that each of the $\easqzeta$ are well-estimated. If this is not true, then we need to augment the error on $\hat{A^2}$ by an extra term which describes the amount of scatter in the residuals. This corresponds to accounting for a non-unity reduced-$\chi^2$ of the WLSQ fit which determines $\hat{A^2}$. Thus our final estimate for the variance of $\hat{A^2}$ is
\begin{eqnarray}
\label{eq:a2esterr_final}
\nonumber \sigma_{\hat{A^2}}^2 & = &\frac{1}{(N_{\rm pairs} - 1)} \frac{\sum_m\left(\left[A^2_{m} \zeta_m\right] - \hat{A^2} \zeta_m\right)^2   /    \sigma^2_{A^2_{m} \zeta_m}  }{ \sum_m \zeta_m^2 /  \sigma^2_{A^2_{m} \zeta_m} } \\
& = & \frac{1}{\left(N_{\rm pairs} - 1\right)}   \frac{\sum \left(A_m^2 - \hat{A^2} \right)^2 / \sigma^2_{A_m}}{\sum 1/\sigma^2_{A_m}}
\end{eqnarray}
which is just the weighted estimate of the variance of $\hat{A^2}$. If $\hat{A^2}$ is significantly larger than $\sigma_{\hat{A^2}}$, then a detection of the GWB has been made. This algorithm has been implemented as a \textsc{tempo2} plugin\footnote{The C codes for this plugin and others from this paper are available at \emph{http://www.atnf.csiro.au/research/pulsar/tempo2/index.php?n= Main.Plugins}}.

\section{Results} \label{sec:res}

\begin{figure}
\includegraphics[width=0.33\textwidth, angle = 270]{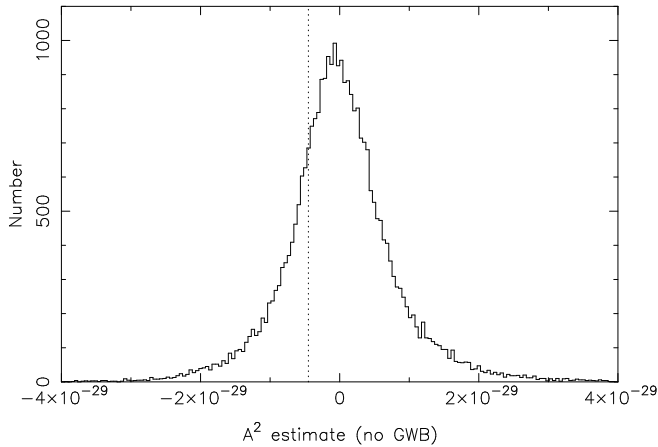} 
\vspace{5pt}
\caption{
The histogram shows the distribution of $\hat{A^2}$ for simulations of the Verbiest et al. (2008, 2009) residuals with no GWB present. The thin dotted line shows the value of $\hat{A^2}$ obtained from the observations. The estimates to the right of the dotted line include 76 per cent of the simulation results. All physical GWBs have $A^2 > 0$.
} \label{fig:distnull}
\end{figure}

From the Verbiest et al. (2009) observations we estimate the squared GWB amplitude to be $\hat{A^2} = -4.5\e{-30}$, with an uncertainty $\sigma_{\hat{A^2}} = 9.1\e{-30}$. Our result is consistent with the null hypothesis, that there is no GWB present. Although the estimate is negative and therefore would lead to an unphysical GWB, it is not improbable because the standard deviation is a factor of 2 larger than the magnitude of the mean. We simulated many realisations of the Verbiest et al. (2009) observations, including the uncertainty given by the ToA error bars and a random process consistent with the low-frequency spectrum of the residuals but no GWB signal. These simulations showed that our estimate is consistent with the null hypothesis with 76 per cent confidence. This result is shown as a histogram in Figure \ref{fig:distnull}. At first, one might think that this histogram could be used to provide a 95 per cent confidence upper bound on the GWB amplitude. However, as discussed further below, any limit thus obtained would not take account of self-noise \citep{jhlm05} due to the GWB-induced perturbations at the pulsar. 

\begin{figure}
\includegraphics[width=0.33\textwidth, angle = 270]{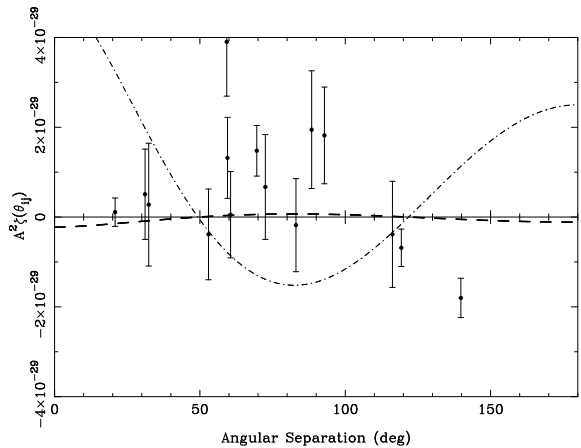} 
\vspace{5pt}
\caption{
The 15 most precise estimates of $\asqzeta$ in the Verbiest et al. (2008, 2009) observations (points with error bars), the best-fit value of $\hat{A^2}\zeta = -4.5\e{-30} \times \zeta$ (dashed curve) and the signal expected from a strong GWB with $A^2 = 1\e{-28}$ (dot-dashed curve).
} \label{fig:a2zeta}
\end{figure}

In Figure \ref{fig:a2zeta}, we plot the 15 estimates of $\asqzeta$ with the smallest uncertainty. It is clear from this figure that the current noise levels are larger than $4.5\e{-30}$ and our result is consistent with the null hypothesis. One might infer from the dot-dashed curve for $A^2 = 1\e{-28}$ that such a large GWB signal is ruled out by the observations. These observations may indeed rule out such a GWB signal, but if $A^2$ were actually $1\e{-28}$ the noise levels on each $\asqzeta$, which provide the upper bound, would be much higher. As the noise levels come from the power spectrum of the residuals of each pulsar, obtaining an upper bound using the noise levels is equivalent to obtaining an upper bound directly from the individual power spectra and ignoring the cross correlations. We will not pursue this bounding technique further in this paper as we are concentrating on the subject of detection.

\section{Discussion} \label{sec:disc}
\begin{figure}
\includegraphics[width=0.33\textwidth, angle = 270]{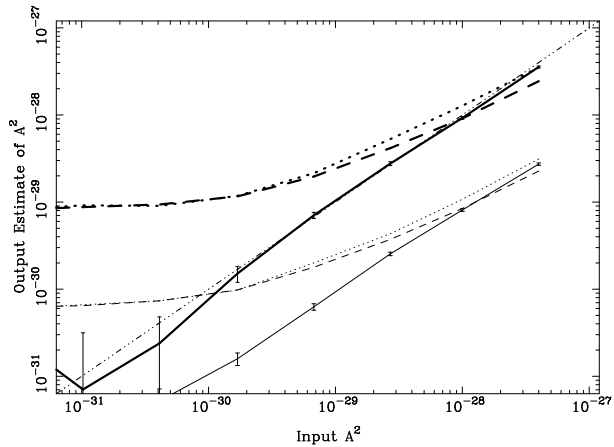}
\vspace{5pt}
\caption{
Average $\hat{A^2}$ as a function of input GWB $A^2$ for the Verbiest et al. (2008, 2009) residuals. The ordinate gives the average output $\hat{A^2}$ from our detection algorithm. The triple-dot-dashed line indicates points where the input $A^2$ is equal to the output $\hat{A^2}$. We have considered 2 cases: performing the full detection procedure (thick lines) and the uncalibrated detection procedure which uses $\gamma_{ij}(f) \equiv 1$ (thin lines). In both cases we have averaged over 1400 realisations for each input $A^2$, and estimated the average output $\hat{A^2}$ (solid lines), where the error bars give the error in the mean of $\hat{A^2}$. The dashed lines give the square root of the average of $\sigma^2_{\hat{A^2}}$ in each case, and are in good agreement with the sample standard deviations over the amplitude range of interest (dotted lines).
} 
\label{fig:AinVsAout}
\end{figure}

\begin{figure}
\includegraphics[width=0.33\textwidth, angle = 270]{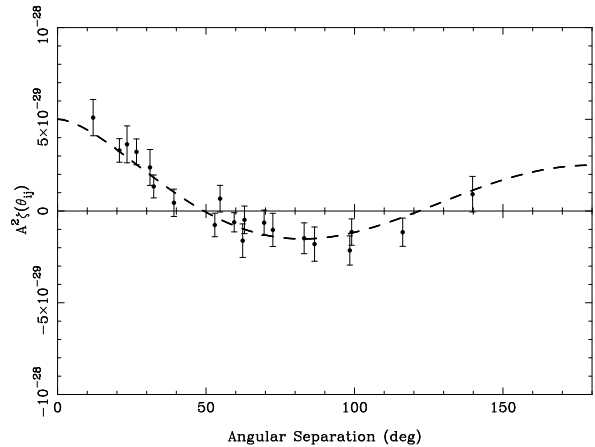}
\vspace{5pt}
\caption{
The expected covariance in simulated residuals which include a GWB component with squared amplitude $A^2=1\e{-28}$. The smooth dashed curve corresponds to the theoretical covariance for an input $A^2=1\e{-28}$. The points correspond to the mean estimates of $\asqzeta$ (see Equations \ref{eq:zeta} and \ref{eq:h_c}) from 200 simulated sets of timing residuals for the 20 PPTA pulsars. The error bars give the uncertainties in these mean estimates. For clarity we only plot the 20 pairs with the smallest rms scatter in their estimates of $\asqzeta$ over the 200 simulations.
} \label{fig:HD2}
\end{figure}

The results of applying this algorithm to the Verbiest et al. (2008, 2009) data are disappointing in the sense that the sensitivity is considerably lower than that calculated in the appendix to Verbiest et al. (2009). We believe the estimated errors to be correct because they are calibrated by simulation, so we ask the question: \emph{Why are the cross power spectra of the GWB lower than expected?} To investigate this we have run a series of simulations\footnote{These simulations use a spread of pulsar distances and synthesise residuals with the same sampling as the Verbiest et al. (2008, 2009) observations. The simulated residuals include white noise consistent with the observed error bars, red noise consistent with the spectral model mentioned in equation (\ref{eq:delpij}) and a signal from a GWB with $\alpha=-2/3$ and with a range of amplitudes between $A^2=6.4\e{-33}$ and $A^2=4\e{-28}$. We did not perform post-Keplerian parameter fits.} with GWB signals of differing amplitudes injected into the observations \citep{hjl+09}. The results are shown in Figure \ref{fig:AinVsAout}. The mean values of the derived $\hat{A^2}$ are plotted as solid lines connecting error bars (which indicate the uncertainty in the mean) for two cases: (1) the algorithm including correction with the $\gamma_{ij}$ calibration factors (thick solid line); and (2) the algorithm with $\gamma_{ij} \equiv 1$ (thin solid line). These results show that our method returns a GWB amplitude estimate $\hat{A}_{\rm out}^2$ such that, on average, $\hat{A}_{\rm out}^2 = A_{\rm in}^2$. Figure \ref{fig:HD2} shows that this GWB signal is at the correct level on average in every pulsar pair. The difference between the thick solid line and the thin solid line in Figure \ref{fig:AinVsAout} indicates that the GWB power is reduced by a factor of $\sim$12 because of the pulsar parameter fitting.

We can estimate the amount of GWB signal lost in estimation of different timing parameters by calculating the weighted average calibration factor in the lowest frequency channel of each pulsar pair. While this will be at a different frequency for each pair, it nevertheless provides a straightforward figure of merit for comparing the effect of fitting different timing model parameters. For the full \textsc{tempo2} fit acting on the Verbiest et al. (2008, 2009) residuals, we find $\overline{\gamma_{ij}}(f=1/\tol) = 0.0790\pm0.0002$, which represents an average loss of $0.0790^{-1}=12.7$ in GWB signal at $f=1/\tol$. This explains the large decrease in sensitivity of our method compared to that presented in the appendix of \citet{vbc+09}, which did not fully account for the effect of pulsar parameter estimation on the GWB signal. In Table \ref{tbl:fit} we show the weighted average calibration factor at $f = 1 / \tol$ when fitting for different parameters in the timing model. The estimation of the pulsar position and parallax have little effect on $\overline{\gamma_{ij}}(f=1/\tol)$ since $\tol$ is a few times greater than 1\,yr for most of our pulsar pairs, and so are not shown in Table \ref{tbl:fit}. This table indicates that one can almost determine the complete effect of fitting on the GWB sensitivity by only including fits for the spin frequency, its derivative and the arbitrary phase offsets between different observing systems. Additionally, while the spin frequency derivative fit only significantly affects the power in the lowest frequency channel, the arbitrary phase offsets affect the power in the lowest few channels which can significantly affect our estimate of $A^2$.

The dashed lines in Figure \ref{fig:AinVsAout} show that for GWB amplitudes around $A^2 = 5\e{-30}$, the average uncertainty on $\hat{A^2}$ is double the average uncertainty when there is no input GWB. This extra contribution to the uncertainty comes from the effect of the GWs passing near the pulsar, which we refer to as the ``self-noise'' of the GWB. For larger values of $A^2$, the uncertainty on $\hat{A^2}$ is dominated by the GWB self-noise as discussed in \citet{jhlm05}. This provides a limitation on the confidence with which we can place an upper bound on the amplitude of the GWB. Because of the self-noise of the GWB, we can obtain at best an 80 per cent confidence upper bound on the GWB amplitude; we can never obtain a 95 per cent confidence bound with our current time series and weighting scheme. Furthermore, any limit obtained thus would be considerably worse than one obtained through other methods, such as direct power estimation, because of the huge variation in noise levels amongst our pulsars\footnote{The \citet{jhv+06} limit method requires that the timing residuals of each pulsar be white, so it cannot be used on these observations. The method presented in \citet{2009MNRAS.395.1005V} could be applied to these observations, but this would require a large amount of computation time and any limit obtained would be difficult to confirm via Monte Carlo simulation.}.

\begin{table}
  \begin{tabular}{@{}cccc@{}}
  \hline
    Timing & Weighted & Uncertainty & Sensitivity \\
    Model & mean of & in Weighted & Loss \\
    Parameters & $\gamma_{ij}(f=1/\tol)$ & Mean & Factor \\
 \hline
 $\nu$, $\dot{\nu}$ &  0.1716 & 0.0003 & 5.83 \\
 $\nu$, $\dot{\nu}$, JUMP & 0.0796 & 0.0002 & 12.6 \\ 
 ALL & 0.0790 & 0.0002 & 12.7 \\
\hline
\end{tabular}
\vspace{-0.0cm}
  \caption{
The effect of fitting different combinations of timing model parameters on the GWB signal in the lowest frequency channel. Values in the 4th column are the inverse of values in the 2nd column. The symbols are: $\nu$ (pulse frequency); $\dot{\nu}$ (pulse frequency derivative); ``JUMP'' (arbitrary phase offsets between different observing systems were removed from all pulsars); ``ALL'' (all timing model parameters were fit).
  } \label{tbl:fit}
\end{table}

We confirm the accuracy of the measured uncertainty on each estimate of $\asqzeta$ using the reduced-$\chi^2$ of the WLSQ fit that determines $\hat{A^2}$. The reduced-$\chi^2$ of this fit is
\begin{equation}
\label{ }
\chi^2_{\rm r} = \frac{1}{(N_{\rm pairs} - 1)}\sum_k\frac{\left(\left[A^2_k \zeta_k\right] - \hat{A^2} \zeta_k\right)^2}{\sigma^2_{A^2_k \zeta_k}}
\end{equation}
which has a value of 1.3 for the Verbiest et al. (2008, 2009) residuals, indicating that the uncertainty estimates $\sigma_{A_k^2}$ are consistent with the rms variation of the estimates $A_k^2$. We obtain an independent estimate of the accuracy of the measured errors by making use of the information contained in the imaginary part of the cross power spectrum, which we denote $\rmn{Imag}\left[X_{ij}(f)\right]$. We calculate $\rmn{Imag}\left[\asqzeta\right]$ by evaluating equation (\ref{eq:a2zeta}) with $\rmn{Imag}\left[X_{ij}(f)\right]$ in place of $\rmn{Real}\left[X_{ij}(f)\right]$. We then process $\rmn{Imag}\left[\asqzeta\right]$ in exactly the same way as the real part is processed. Since correlation coefficients are real, we expect that $\rmn{Imag}\left[\asqzeta\right]$ will contain no correlated signal. This means that we can calculate the analogue of the reduced-$\chi^2$ using $\rmn{Imag}\left[\asqzeta\right]$:
\begin{equation}
\label{}
\chi^2_{\rm r, im} = \frac{1}{(N_{\rm pairs} - 1)}\sum_k\frac{\left(\rmn{Imag}\left[A^2_k \zeta_k\right]\right)^2}{\sigma^2_{A^2 _k\zeta_k}} \; .
\end{equation}
Similar to the reduced-$\chi^2$, if the errors on $\asqzeta$ are well-estimated then this quantity should be near unity. For the Verbiest et al. (2008, 2009) residuals, we find $\chi^2_{\rm r, im} = 0.87$, indicating that the errors are well-estimated.

Although both $\chi^2_{\rm r}$ and $\chi^2_{\rm r, im}$ show that the uncertainties $\sigma_{A^2_k}$ are reliable on average, these uncertainties come from power spectral estimates so they are random variables. We estimated the sensitivity of $\hat{A^2}$ to variations in $\sigma_{A_k^2}$ by multiplying each $\sigma_{A_k^2}$ by a random factor, distributed as the square root of the product of two $\chi^2$ random variables with two degrees of freedom. This is the expected distribution for each $\sigma_{A_k^2}$.
We found that $\sigma_{\hat{A^2}}$ increased by a factor of 1.6, indicating that the use of incorrect $\easqzeta$ estimates degrades the sensitivity of the $\hat{A^2}$ measurement by only a factor of 1.6.

However, the $\asqzeta$ are not Gaussian; rather they come from the sum of two pairwise products of independent Gaussian variables and thus have a two-sided exponential distribution which is reflected in Figure \ref{fig:distnull}. This means that the maximum likelihood estimator for $A^2$ is not a WLSQ estimator but a weighted least absolute deviation (LAD) fit \citep[see, e.g.,][]{2006Cox}. We tested both weighted and unweighted LAD fits and found that the results for WLSQ and unweighted LAD fits were very similar, while the weighted LAD fit introduced a small bias in the mean. These results are shown in Table \ref{tbl:est}. We suspect that the bias occurs because any LAD fit includes a `dead-zone' feature, where a range of parameter estimates give the same minimum absolute deviation. This dead zone is negligible when the number of estimates is large, but can be significant otherwise. Since our $A^2$ estimates are dominated by a small number of $A^2_k$ measurements and the results of the different estimators are similar, we chose the more standard WLSQ fit in calculating $\hat{A^2}$. Although the WLSQ estimator is not maximum likelihood, it is apparently more robust in our particular case.

\begin{table}
  \caption{
The results from estimating $A^2$ with different estimators averaged over $10^5$ simulations of realistic residuals including a GWB with $A^2 = 1\e{-30}$.} \label{tbl:est}
  \begin{tabular}{@{}cccc@{}}
  \hline
    Estimator & Mean $\hat{A^2}$ & Error in Mean & rms of $\hat{A^2}$\\
 & ($\e{-30}$) & $\hat{A^2}$ ($\e{-30}$) & ($\e{-30}$) \\
 \hline
WLSQ [our method] & 0.99 & 0.038 & 12 \\
Unweighted LAD & 1.0 & 0.038 & 12 \\
Weighted LAD & 0.84 & 0.041 & 13 \\
\hline
\end{tabular}
\vspace{-0.0cm}
\end{table}

Estimation of $A^2$ is also largely independent of changes to the method of spectral analysis. We experimented with reducing the white noise in the residuals by smoothing each time series over a 60-day period before commencing the spectral analysis. We also tested interpolation using a constrained cubic spline of each smoothed time series onto a 14-day grid common to all pulsars before the spectral analysis. The results of these different approaches are given in Table \ref{tbl:specanal}. Since there was no statistically significant difference between the different approaches, for simplicity we elected not to smooth or interpolate the residuals.

\begin{table}
  \caption{
The results from the Verbiest et al. (2008, 2009) observations using different methods of spectral analysis of the timing residuals.
  } \label{tbl:specanal}
  \begin{tabular}{@{}ccc@{}}
  \hline
    Processing & $\hat{A^2}$ & $\sigma_{\hat{A^2}}$\\
    Performed & ($\e{-30}$) & ($\e{-30}$)\\
 \hline
Smoothing \& Interpolation & $3.0$ & $10$ \\
Smoothing only & $-7.8$  & $10$ \\
No smoothing [our method] & $-4.5$ & $9.1$ \\
\hline
\end{tabular}
\vspace{-0.0cm}
\end{table}

\subsection{Treatment of large GWB signals}

For their detection statistic, \citet{jhlm05} calculate the normalised cross correlation between the timing residuals of each pulsar pair. They optimise the S/N ratio using a filter designed to whiten the residuals before correlation. For a simulation of the 20 PPTA pulsars, this approach increased the maximum achievable detection significance for a GWB from 3$\sigma$ to 13$\sigma$. However, their filter cannot be applied to real pulsar timing observations without modification. We investigated the effect of such a filter by performing simulations of the Verbiest et al. (2008, 2009) residuals where each simulation included a signal from a GWB with $A \gtrsim 3\e{-15}$. In the frequency domain, the filter takes the form of a weighting factor, so we optimised this weighting factor to match the large input GWB amplitude. We found that this method did not improve the S/N ratio, and we traced this under-performance to the problem of spectral leakage from the lowest frequencies to the higher frequencies. We found that the first few cross-spectral estimates, which make the largest contribution to our detection statistic, were all more than 90 per cent correlated with the lowest spectral estimate (i.e., at frequency $f = 1/\tol$), meaning that re-weighting cannot change the overall S/N ratio. The spectral leakage is particularly significant because of the irregular sampling and variable ToA uncertainties in these observations. We expect that an improved spectral analysis technique \citep[e.g.,][]{2010Coles} will eliminate the spectral leakage and enable us to take advantage of more degrees of freedom\footnote{In contrast to \citet{vbc+09} which states that quadratic fitting removes one degree of freedom from the power spectrum of each pulsar's residuals, our analysis has shown that quadratic fitting does not affect the number of degrees of freedom in the lowest few frequency channels of each power spectrum.} when the GWB signal is larger than the noise.

\subsection{Fitting timing models over different data spans}

\begin{figure}
\includegraphics[width=0.33\textwidth, angle = 270]{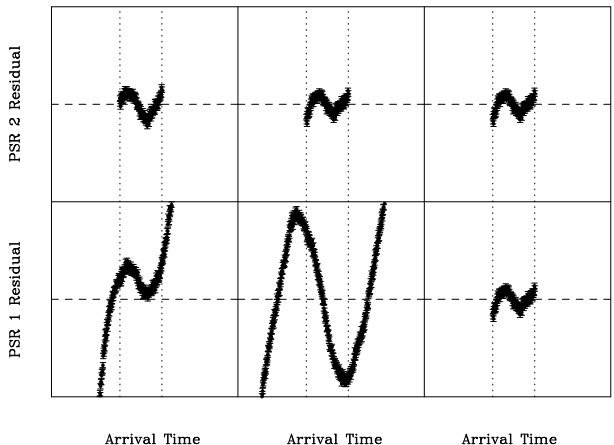}
\vspace{5pt}
\caption{
The effect of fitting a timing model over different data spans. The time series in the upper three panels are 5 years long, the time series in the lower three panels are 15 years long (to keep the y-axis scaling consistent, the plotting window has truncated the longer time series in the first 2 panels, and the bottom right panel only includes the overlapping data). The vertical dotted lines indicate the overlapping timing residuals for these time series. We added the same large signal to both time series and the time series are identical in the overlapping region (left panels). After fitting the timing model (middle panels), this signal is no longer correlated between the two time series. The correlation is restored by performing a WLSQ fit of a quadratic function in the overlapping region of the two time series (right panels).
} \label{fig:fit}
\end{figure}

The time series we consider in this paper have widely varying time spans, which has not been a feature of most PTA analyses to date. As part of the pulsar parameter estimation, we fit for the pulse period and its derivative over the full duration of each time series. Originally, we then computed the cross power spectra from the overlapping portion of residuals of each pulsar pair with no further processing. However, upon simulating this procedure, we found that the lowest frequencies in the cross power spectra were biased whenever $\tobs > \tol$. This bias took the form of a significantly non-zero imaginary part in the cross power spectrum. Also, we found that much of the correlated signal at low frequencies was removed, as shown in Figure \ref{fig:fit}. We were unable to eliminate these effects unless we performed a WLSQ fit of a quadratic function for each time series over the overlapping time range. This restores the correlation in the GWB signal between different pulsars (right panels of Figure \ref{fig:fit}). This additional WLSQ fit will introduce a new bias because of removing some of the GWB signal at $f=1/\tol$, but this new bias is easily corrected with the calibration factors $\gamma_{ij}(f)$. However, there is an additional loss of 10 per cent of the GWB signal in the Verbiest et al. (2008, 2009) observations because of this extra WLSQ fit.

\subsection{Correlated signals in the timing residuals} \label{sec:corrsig}

The GWB analysis is complicated by the unknown effects of other correlated signals in the timing residuals. Instabilities in TT and errors in the Solar-System ephemeris both produce signals which are correlated between different pulsars. We estimated the effect of these uncertainties by using an updated timescale and the most recent Solar-System ephemeris.

\begin{table}
  \caption{
The results from using updated realisations of TT and the Solar-System ephemeris. The last column gives the change in the value of $\hat{A^2}$ with respect to processing the observations with TT(TAI) and DE405, the realisations used for the Verbiest et al. (2008, 2009) residuals.
  } \label{tbl:ttephem}
  \begin{tabular}{@{}ccccc@{}}
  \hline
    Realisation 	 & Solar 		& 			& 			& \hspace{-3mm} Change \\
    of Terrestrial	& System 		& $\hat{A^2}$	& \hspace{-3mm} $\sigma_{\hat{A^2}}$	& \hspace{-3mm} in $\hat{A^2}$ \\
     Time		& Ephemeris	& $(\times 10^{-30})$	& \hspace{-3mm} $(\times 10^{-30})$	&  \hspace{-3mm} $(\times 10^{-30})$\\
 \hline
TT(TAI) 	& DE405 	& $-4.5$	& \hspace{-3mm} $9.1$	& \hspace{-3mm} 0.0  \\
TT(TAI) 	& DE421 	& $-2.3$	& \hspace{-3mm} $9.4$	& \hspace{-3mm} $2.2$  \\
TT(BIPM2010) 	& DE405 	& $-3.7$	& \hspace{-3mm} $8.7$	& \hspace{-3mm} $0.8$  \\
\hline
\end{tabular}
\vspace{-0.0cm}
\end{table}

Instabilities in TT produce a positive cross correlation independent of angular separation. Any estimate of the clock error will thus be correlated with the estimate of the GWB amplitude. Had we made a significant detection of the GWB, this would have to be accounted for. To estimate the importance of possible clock instabilities, we processed the Verbiest et al. (2008, 2009) observations using the version of TT released by BIPM in 2010 \citep[see, e.g.,][]{2003Petit}. This post-corrected timescale has revealed statistically significant inaccuracies in TT(TAI). The results are shown in Table \ref{tbl:ttephem}. While the change of clock reference only changes our estimated GWB level by nine per cent of the uncertainty, the absolute change ($0.8\e{-30}$) is at a significant level for some predictions of the GWB \citep{jb03,svc08}. This implies that such instabilities in TT must be accounted for when analysing future data sets.

The results from using the newest Solar-System ephemeris DE421 \citep{2009DE421} are given in Table \ref{tbl:ttephem}. While there have been some improvements in this ephemeris version compared to DE405, most of the changes are absorbed by the pulsar parameter fit. The estimated GWB level has changed by 24 per cent of the uncertainty. If we assume DE421 is correct, then the use of DE405 is similar to introducing a spurious GWB signal with $A=1.5\e{-15}$, a signal which is undetectable in most time series from the Verbiest et al. (2008, 2009) observations. However, future observations will need to account for the effects of inaccuracies in the Solar-System ephemeris.

\subsection{Contribution of different pulsars to $\hat{A^2}$}

It is difficult to determine the exact contributions to the weighting of each pulsar pair when using error bars derived from Monte Carlo simulations. The dominant effect is the size of $\tol$. For a GWB caused by SMBHBs, the weighting factor increases approximately as $\tol^{4.3}$. A higher noise level in the residuals of each pulsar in the pair will decrease the weight of that pair approximately linearly. The angle subtended at the observer by the pair of pulsars $\theta_{ij}$ can be important if $\theta_{ij}$ is near the zeroes of the function plotted in Figure \ref{fig:HD}.

\begin{figure}
\includegraphics[width=0.3\textwidth, angle = 270]{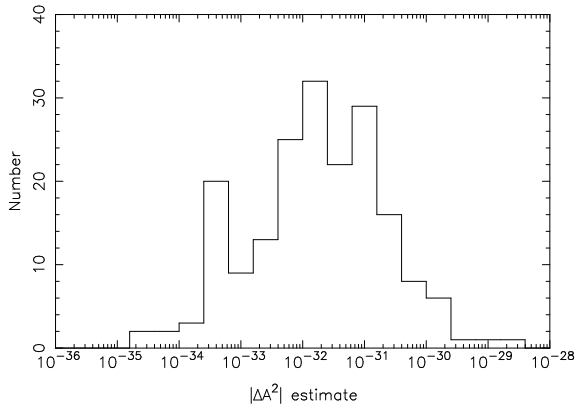}
\vspace{5pt}
\caption{
The effect on $\hat{A^2}$ of the removal of different pulsar pairs, as measured by $|\Delta\hat{A^2}|$. Almost all pulsar pairs have no significant effect on the value of $\hat{A^2}$ obtained from the Verbiest et al. (2008, 2009) residuals.
} \label{fig:histdelA2}
\end{figure}

To determine which pulsars contribute the most to our estimate of the GWB, we perform the WLSQ fit described by Equations (\ref{eq:a2est}) and (\ref{eq:a2esterr}) to only 189 of the possible 190 $\asqzeta$ estimates. By varying which estimate of $\asqzeta$ is removed, we can find the pulsar pairs which have the greatest influence over the measurement of $\hat{A^2}$ in these residuals. This is performed by finding $\Delta\hat{A^2}$ for each pair of pulsars, which is the measured $\hat{A^2}$ from all pulsar pairs minus the value of $\hat{A^2}$ when \emph{not} including the given pulsar pair. Those pairs with the largest contribution to this measure are given in Table \ref{tbl:psrpairs}, and a histogram of the absolute value $\left|\Delta\hat{A^2}\right|$ for all pulsar pairs is provided in Figure \ref{fig:histdelA2}.

\begin{table}
  \caption{
  The nine pulsar pairs whose absence from the array changes the measurement of $\hat{A^2}$ from the Verbiest et al. (2008, 2009) residuals by more than $1\e{-30}$. The first column contains the names of the pulsars in the pair, the second column lists values of $\Delta\hat{A^2}$, and the third column gives the change as a percentage of the value of $\hat{A^2}$ derived when using all our data.
  } \label{tbl:psrpairs}
  \begin{tabular}{@{}ccc@{}}
  \hline
    Removed Pulsar Pair & $\Delta\hat{A^2}$ ($\e{-30}$) & Percentage change \\
\hline
J1713+0747, J1744-1134 &  18.0  & -400\% \\
J2124-3358, J2145-0750 &  2.32  & -52\% \\
J1730-2304, J1744-1134 &  2.10  & -47\% \\
J0711-6830, J2145-0750 &  1.26  & -28\% \\
\hline
J0437-4715, J1909-3744 &  -1.07 & 24\% \\
J0437-4715, J2129-5721 &  -1.36 & 30\% \\
J0437-4715, J2145-0750 &  -1.41 & 31\% \\
J1713+0747, J2145-0750 &  -3.97 & 88\% \\
J0437-4715, J1713+0747 &  -7.15 & 159\% \\
\hline
\end{tabular}
\vspace{-0.0cm}
\end{table}

This analysis shows that the measurement of $\hat{A^2}$ is determined by only a few pulsar pairs. This severely reduces the number of degrees of freedom when detecting the GWB, and thus decreases the maximum attainable detection confidence \citep[see][]{jhlm05} because it reduces our ability to average out the self-noise in the residuals caused by the GWB signal at each pulsar. Observing more strong pulsars is essential to increasing the number of degrees of freedom in order to detect the GWB with reasonable confidence. This is further endorsement of the International Pulsar Timing Array concept \citep{2010CQGra..27h4013H}, but contrary to a suggested strategy for detection of individual GW sources \citep{2010arXiv1005.5163B}. This is a fundamental difference between the single GW source detection problem and the GWB detection problem.

\section{Conclusions} \label{sec:conc}

In implementing a GWB detection algorithm along the lines originally proposed by \citet{jhlm05} we have confronted a number of issues which must be addressed when using real observations. We find that in practice the S/N ratio can be reduced by a factor of $\sim$12 compared with the ideal situation discussed by \citet{vbc+09} because of the fitting of a timing model to form the residuals. In particular, almost all of the signal loss is caused by the fitting of a quadratic term and arbitrary phase offsets between different observing systems. We also find that it will be important to estimate and correct both clock errors and ephemeris errors when attempting to detect the GWB at a level less than $A = 2\e{-15}$. As pointed out by \citet{jhlm05}, prewhitening will be required to obtain detection significance larger than $3\sigma$. We find that this cannot be done without solving the problem of spectral leakage due to irregular sampling and variable ToA uncertainties.

Fortunately, there are encouraging indications that many of these problems can be solved. Recent work \citep{2010Hobbs, 2010ApJ...720L.201C} shows that clock errors and ephemeris errors can be estimated and removed. These errors are at a level which would disrupt the GWB signal in pulsar timing observations in the near future, and could even impact the analysis of a modified version of the Verbiest et al. (2008, 2009) observations which did not include arbitrary phase offsets between observing systems. As systems with more sensitivity become available, the clock and ephemeris communities will improve their data sets. It appears possible to improve the process of fitting a timing model and also to improve the spectral leakage using the algorithm discussed by \citet{2010Coles}. It has proved possible to calibrate most of the phase discontinuities between different observing systems in the PPTA observations and this alone can improve the S/N ratio by a factor of two.

We have not discussed DM variations, but it is likely that some of the low frequency noise in our residuals is due to such interstellar propagation effects. Certainly as the various PTA data sets improve it will be essential to estimate and remove any frequency-dependent effects.

Our analysis shows that, although the Verbiest et al. (2008, 2009) data set contains observations of 20 pulsars spanning many years, only a few of the pulsars in this data set contribute significantly to detecting the GWB, thereby reducing our detection confidence. It is uncertain whether this will be the case for the most recent observations from the PPTA. Observations of a larger sample of pulsars with precise ToA measurements will help to overcome this problem.

\section{Acknowledgments}

The authors would like to thank Yuri Levin for commenting on a draft version of this paper, and Xavier Siemens, Larry Price and Kejia Lee for useful discussions.

This work is undertaken as part of the Parkes Pulsar Timing Array project which was initiated with support from RNM's Australian Research Council Federation Fellowship (\#FF0348478). The Parkes radio telescope is part of the Australia Telescope, which is funded by the Commonwealth of Australia for operation as a National Facility managed by the Commonwealth Scientific and Industrial Research Organisation (CSIRO). DRBY is funded by an APA and the CSIRO OCE PhD scholarship program. GH is the recipient of an Australian Research Council QEII Fellowship (\#DP0878388). JPWV is supported by the European Union under Marie Curie Intra-European Fellowship 236394. MB acknowledges support from ARC grant DP0985272.

\bibliographystyle{mn2e}
\bibliography{journals,psrrefs,mybibliography,modrefs,crossrefs}

\hyphenation{Post-Script Sprin-ger}
\begin{thebibliography}{}

\bibitem[\protect\citeauthoryear{{Anholm}, {Ballmer}, {Creighton}, {Price} \&
  {Siemens}}{{Anholm} et~al.}{2009}]{2009PhRvD..79h4030A}
{Anholm} M.,  {Ballmer} S.,  {Creighton} J.~D.~E.,  {Price} L.~R.,    {Siemens}
  X.,  2009, \prd, 79, 084030

\bibitem[\protect\citeauthoryear{{Bailes}}{{Bailes}}{2003}]{2003ASPC..302...57%
B}
{Bailes} M.,  2003, in {M.~Bailes, D.~J.~Nice, \& S.~E.~Thorsett} ed., Radio
  Pulsars Vol.~302 of Astronomical Society of the Pacific Conference Series.
p.~57

\bibitem[\protect\citeauthoryear{Blandford, Narayan \& Romani}{Blandford
  et~al.}{1984}]{bnr84}
Blandford R.~D.,  Narayan R.,    Romani R.~W.,  1984, J. Astrophys. Astr., 5,
  369

\bibitem[\protect\citeauthoryear{{Burt}, {Lommen} \& {Finn}}{{Burt}
  et~al.}{2010}]{2010arXiv1005.5163B}
{Burt} B.~J.,  {Lommen} A.~N.,    {Finn} L.~S.,  2010, arXiv: 1005.5163

\bibitem[\protect\citeauthoryear{{Champion}, {Hobbs}, {Manchester} \& {et
  al.}}{{Champion} et~al.}{2010}]{2010ApJ...720L.201C}
{Champion} D.~J.,  {Hobbs} G.~B.,  {Manchester} R.~N.,    {et al.} 2010, \apjl,
  720, L201

\bibitem[\protect\citeauthoryear{{Coles}, {Hobbs}, {Champion} \& {et
  al.}}{{Coles} et~al.}{2010}]{2010Coles}
{Coles} W.~A.,  {Hobbs} G.~B.,  {Champion} D.~J.,    {et al.} 2010, submitted
  to MNRAS

\bibitem[\protect\citeauthoryear{{Corbin} \& {Cornish}}{{Corbin} \&
  {Cornish}}{2010}]{2010arXiv1008.1782C}
{Corbin} V.,  {Cornish} N.~J.,  2010, arXiv:astro-ph/1008.1782

\bibitem[\protect\citeauthoryear{{Cox}}{{Cox}}{2006}]{2006Cox}
{Cox} D.~R.,  2006, Principles of Statistical Inference.
Cambridge University Press

\bibitem[\protect\citeauthoryear{{Damour} \& {Vilenkin}}{{Damour} \&
  {Vilenkin}}{2005}]{dv05}
{Damour} T.,  {Vilenkin} A.,  2005, Phys. Rev. D, 71, 063510

\bibitem[\protect\citeauthoryear{Detweiler}{Detweiler}{1979}]{det79}
Detweiler S.,  1979, \apj, 234, 1100

\bibitem[\protect\citeauthoryear{{Edwards}, {Hobbs} \& {Manchester}}{{Edwards}
  et~al.}{2006}]{ehm06}
{Edwards} R.~T.,  {Hobbs} G.~B.,    {Manchester} R.~N.,  2006, \mnras, 372,
  1549

\bibitem[\protect\citeauthoryear{{Folkner}, {Williams} \& {Boggs}}{{Folkner}
  et~al.}{2009}]{2009DE421}
{Folkner} W.~M.,  {Williams} J.~G.,    {Boggs} D.~H.,  2009, IPN Progress
  Report 42-178 (Pasadena, CA: NASA Jet Propulsion Laboratory

\bibitem[\protect\citeauthoryear{{Hellings}}{{Hellings}}{1989}]{1989NASCP3046.%
..93H}
{Hellings} R.~W.,  1989, in {R.~W.~Hellings} ed., NASA Conference Publication
  Vol.~3046 of NASA Conference Publication, {Pulsar timing and gravitational
  waves}.
pp 93--97

\bibitem[\protect\citeauthoryear{Hellings \& Downs}{Hellings \&
  Downs}{1983}]{hd83}
Hellings R.~W.,  Downs G.~S.,  1983, \apjl, 265, L39

\bibitem[\protect\citeauthoryear{{Hobbs}, {Archibald}, {Arzoumanian} \& {et
  al.}}{{Hobbs} et~al.}{2010a}]{2010CQGra..27h4013H}
{Hobbs} G.,  {Archibald} A.,  {Arzoumanian} Z.,    {et al.} 2010a, Classical and
  Quantum Gravity, 27, 084013

\bibitem[\protect\citeauthoryear{{Hobbs}, {Coles}, {Manchester} \&
  {Chen}}{{Hobbs} et~al.}{2010b}]{2010arXiv1011.5285H}
{Hobbs} G.,  {Coles} W.,  {Manchester} R.,    {Chen} D.,  2010b,
  arXiv:astro-ph/1011.5285

\bibitem[\protect\citeauthoryear{{Hobbs}, {Jenet}, {Lee} \& {et al.}}{{Hobbs}
  et~al.}{2009}]{hjl+09}
{Hobbs} G.,  {Jenet} F.,  {Lee} K.~J.,    {et al.} 2009, MNRAS, 394, 1945

\bibitem[\protect\citeauthoryear{{Hobbs}, {Lyne} \& {Kramer}}{{Hobbs}
  et~al.}{2010c}]{2010MNRAS.402.1027H}
{Hobbs} G.,  {Lyne} A.~G.,    {Kramer} M.,  2010c, \mnras, 402, 1027

\bibitem[\protect\citeauthoryear{{Hobbs}, {Coles}, {Manchester} \& {et
  al.}}{{Hobbs} et~al.}{2010d}]{2010Hobbs}
{Hobbs} G.~B.,  {Coles} W.~A.,  {Manchester} R.~N.,    {et al.} 2010d, submitted
  to MNRAS

\bibitem[\protect\citeauthoryear{{Hobbs}, {Edwards} \& {Manchester}}{{Hobbs}
  et~al.}{2006}]{hem06}
{Hobbs} G.~B.,  {Edwards} R.~T.,    {Manchester} R.~N.,  2006, \mnras, 369, 655

\bibitem[\protect\citeauthoryear{{Hotan}, {van Straten} \&
  {Manchester}}{{Hotan} et~al.}{2004}]{hvm04}
{Hotan} A.~W.,  {van Straten} W.,    {Manchester} R.~N.,  2004, Proc. Astr.
  Soc. Aust., 21, 302

\bibitem[\protect\citeauthoryear{Jaffe \& Backer}{Jaffe \& Backer}{2003}]{jb03}
Jaffe A.~H.,  Backer D.~C.,  2003, ApJ, 583, 616

\bibitem[\protect\citeauthoryear{{Jenet}, {Hobbs}, {Lee} \&
  {Manchester}}{{Jenet} et~al.}{2005}]{jhlm05}
{Jenet} F.~A.,  {Hobbs} G.~B.,  {Lee} K.~J.,    {Manchester} R.~N.,  2005,
  ApJL, 625, L123

\bibitem[\protect\citeauthoryear{{Jenet}, {Hobbs}, {van Straten} \& {et
  al.}}{{Jenet} et~al.}{2006}]{jhv+06}
{Jenet} F.~A.,  {Hobbs} G.~B.,  {van Straten} W.,    {et al.} 2006, ApJ, 653,
  1571

\bibitem[\protect\citeauthoryear{Kaspi, Taylor \& Ryba}{Kaspi
  et~al.}{1994}]{ktr94}
Kaspi V.~M.,  Taylor J.~H.,    Ryba M.,  1994, \apj, 428, 713

\bibitem[\protect\citeauthoryear{{Lee}, {Jenet} \& {Price}}{{Lee}
  et~al.}{2008}]{2008ApJ...685.1304L}
{Lee} K.~J.,  {Jenet} F.~A.,    {Price} R.~H.,  2008, ApJ, 685, 1304

\bibitem[\protect\citeauthoryear{{Lommen}}{{Lommen}}{2002}]{lom02}
{Lommen} A.~N.,  2002, in Becker W.,  Lesch H.,   Tr\"umper J.,  eds,
  WE-Heraeus Seminar on Neutron Stars, Pulsars, and Supernova Remnants
  Max-Plank-Institut f\"ur Extraterrestrische Physik, Garching, pp 114--125

\bibitem[\protect\citeauthoryear{{Lorimer} \& {Kramer}}{{Lorimer} \&
  {Kramer}}{2004}]{2004hpa..book.....L}
{Lorimer} D.~R.,  {Kramer} M.,  2004, {Handbook of Pulsar Astronomy}.
Vol.~4.~Cambridge, UK: Cambridge University Press

\bibitem[\protect\citeauthoryear{Maggiore}{Maggiore}{2000}]{mag00}
Maggiore M.,  2000, \physrep, 331, 283

\bibitem[\protect\citeauthoryear{{Manchester}}{{Manchester}}{2010}]{2010arXiv1%
004.3602M}
{Manchester} R.~N.,  2010, eprint arXiv:astro-ph/1004.3602

\bibitem[\protect\citeauthoryear{{Navarro}}{{Navarro}}{1994}]{1994PhDT........%
12N}
{Navarro} J.,  1994, PhD thesis, California Inst.~of Tech.

\bibitem[\protect\citeauthoryear{{Petit}}{{Petit}}{2003}]{2003Petit}
{Petit} G.,  2003, 35th Annual Precise Time and Time Interval (PTTI) Meeting,
  pp 307 -- 318

\bibitem[\protect\citeauthoryear{Petit \& Tavella}{Petit \&
  Tavella}{1996}]{pt96}
Petit G.,  Tavella P.,  1996, A\&A, 308, 290

\bibitem[\protect\citeauthoryear{{Phinney}}{{Phinney}}{2001}]{2001astro.ph..80%
28P}
{Phinney} E.~S.,  2001, eprint arXiv:astro-ph/0108028

\bibitem[\protect\citeauthoryear{{Romani} \& {Taylor}}{{Romani} \&
  {Taylor}}{1983}]{1983ApJ...265L..35R}
{Romani} R.~W.,  {Taylor} J.~H.,  1983, ApJL, 265, L35

\bibitem[\protect\citeauthoryear{Sazhin}{Sazhin}{1978}]{saz78}
Sazhin M.~V.,  1978, Sov. Astron., 22, 36

\bibitem[\protect\citeauthoryear{Scargle}{Scargle}{1982}]{sca82}
Scargle J.~D.,  1982, ApJ, 263, 835

\bibitem[\protect\citeauthoryear{{Sesana}, {Vecchio} \& {Colacino}}{{Sesana}
  et~al.}{2008}]{svc08}
{Sesana} A.,  {Vecchio} A.,    {Colacino} C.~N.,  2008, MNRAS, 390, 192

\bibitem[\protect\citeauthoryear{{Sesana}, {Vecchio} \& {Volonteri}}{{Sesana}
  et~al.}{2009}]{svv09}
{Sesana} A.,  {Vecchio} A.,    {Volonteri} M.,  2009, MNRAS, 394, 2255

\bibitem[\protect\citeauthoryear{{Shannon} \& {Cordes}}{{Shannon} \&
  {Cordes}}{2010}]{2010ApJ...725.1607S}
{Shannon} R.~M.,  {Cordes} J.~M.,  2010, \apj, 725, 1607

\bibitem[\protect\citeauthoryear{Standish}{Standish}{2004}]{sta04b}
Standish E.~M.,  2004, A\&A, 417, 1165

\bibitem[\protect\citeauthoryear{Taylor}{Taylor}{1992}]{tay92}
Taylor J.~H.,  1992, Phil. Trans. Roy. Soc. A, 341, 117

\bibitem[\protect\citeauthoryear{Taylor \& Weisberg}{Taylor \&
  Weisberg}{1982}]{tw82}
Taylor J.~H.,  Weisberg J.~M.,  1982, \apj, 253, 908

\bibitem[\protect\citeauthoryear{{Thorsett} \& {Dewey}}{{Thorsett} \&
  {Dewey}}{1996}]{1996PhRvD..53.3468T}
{Thorsett} S.~E.,  {Dewey} R.~J.,  1996, Phys Rev D, 53, 3468

\bibitem[\protect\citeauthoryear{{van Haasteren} \& {Levin}}{{van Haasteren} \&
  {Levin}}{2010}]{2010MNRAS.401.2372V}
{van Haasteren} R.,  {Levin} Y.,  2010, \mnras, 401, 2372

\bibitem[\protect\citeauthoryear{{van Haasteren}, {Levin}, {McDonald} \&
  {Lu}}{{van Haasteren} et~al.}{2009}]{2009MNRAS.395.1005V}
{van Haasteren} R.,  {Levin} Y.,  {McDonald} P.,    {Lu} T.,  2009, \mnras,
  395, 1005

\bibitem[\protect\citeauthoryear{{van Straten}}{{van Straten}}{2006}]{van06}
{van Straten} W.,  2006, \apj, 642, 1004

\bibitem[\protect\citeauthoryear{{van Straten}, {Manchester}, {Johnston} \&
  {Reynolds}}{{van Straten} et~al.}{2010}]{2010PASA...27..104V}
{van Straten} W.,  {Manchester} R.~N.,  {Johnston} S.,    {Reynolds} J.~E.,
  2010, Proc. Astr. Soc. Aust., 27, 104

\bibitem[\protect\citeauthoryear{{Verbiest}, {Bailes}, {Bhat} \& {et
  al.}}{{Verbiest} et~al.}{2010}]{2010CQGra..27h4015V}
{Verbiest} J.~P.~W.,  {Bailes} M.,  {Bhat} N.~D.~R.,    {et al.} 2010, Class.
  and Quant. Grav., 27, 084015

\bibitem[\protect\citeauthoryear{{Verbiest}, {Bailes}, {Coles} \& {et
  al.}}{{Verbiest} et~al.}{2009}]{vbc+09}
{Verbiest} J.~P.~W.,  {Bailes} M.,  {Coles} W.~A.,    {et al.} 2009, MNRAS,
  400, 951

\bibitem[\protect\citeauthoryear{{Verbiest}, {Bailes}, {van Straten} \& {et
  al.}}{{Verbiest} et~al.}{2008}]{vbv+08}
{Verbiest} J.~P.~W.,  {Bailes} M.,  {van Straten} W.,    {et al.} 2008, ApJ,
  679, 675

\bibitem[\protect\citeauthoryear{Wyithe \& Loeb}{Wyithe \&
  Loeb}{2003}]{2003ApJ...590..691W}
Wyithe J.~S.~B.,  Loeb A.,  2003, ApJ, 590, 691

\bibitem[\protect\citeauthoryear{{Yardley}, {Hobbs}, {Jenet} \& {et
  al.}}{{Yardley} et~al.}{2010}]{2010MNRAS.407..669Y}
{Yardley} D.~R.~B.,  {Hobbs} G.~B.,  {Jenet} F.~A.,    {et al.} 2010, \mnras,
  407, 669

\bibitem[\protect\citeauthoryear{{You}, {Hobbs}, {Coles} \& {et al.}}{{You}
  et~al.}{2007}]{yhc+07}
{You} X.~P.,  {Hobbs} G.,  {Coles} W.~A.,    {et al.} 2007, \mnras, 378, 493

\bibitem[\protect\citeauthoryear{{Zechmeister} \& {K{\"u}rster}}{{Zechmeister}
  \& {K{\"u}rster}}{2009}]{2009A&A...496..577Z}
{Zechmeister} M.,  {K{\"u}rster} M.,  2009, \aap, 496, 577

\end{thebibliography}

\label{lastpage}

\end{document}